\documentstyle[tighten,preprint,prd,aps,epsfig]{revtex}

\newcommand{\ds}        {\mbox{$\not\hspace{-0.7truemm}\partial$}}
\newcommand{\asigma}    {\left|\sigma\right|}
\newcommand{\adelta}    {\left|\Delta\right|}
\newcommand{\Vec}[1]    {\mbox{\boldmath$#1$}}
\newcommand{\abs}[1]    {\left| #1 \right|}
\newcommand{\ps}        {\psi_i}
\newcommand{\psb}       {\bar{\psi_i}}
\newcommand{\psd}       {{\psi_i}^{\dag}}
\newcommand{\psch}      {\psi_i^c}
\newcommand{\pscb}      {\bar{\psi_i^c}}
\newcommand{\pscd}      {{\psi_i^c}^{\dag}}
\newcommand{\aap}       {a_{ pi}}
\newcommand{\aam}       {a_{-pi}}
\newcommand{\adp}       {a_{ pi}^{\dag}}
\newcommand{\adm}       {a_{-pi}^{\dag}}
\newcommand{\bbp}       {b_{ pi}}
\newcommand{\bbm}       {b_{-pi}}
\newcommand{\bdp}       {b_{ pi}^{\dag}}
\newcommand{\bdm}       {b_{-pi}^{\dag}}
\newcommand{\vac}       {|0\rangle}
\newcommand{\vacs}      {|\sigma\rangle}
\newcommand{\vacd}      {|\Delta\rangle}

\begin{document}

\draft

%%%%%%%%%%^%%%%%%%%%^%%%%%%%%%^%%%%%%%%%^%%%%%%%%%^%%%%%%%%% Title Page
\title{ Crystalline ground state in chiral Gross-Neveu and Cooper pair
 models at finite densities }

\author{ Katsumi Ohwa\footnote{k\_ohwa@comp.metro-u.ac.jp} }

\address{ Department of Physics, Tokyo Metropolitan University,\\
             1-1 Minami-Osawa, Hachioji, Tokyo, 192-0397, Japan }

\date{\today}

\maketitle

\begin{abstract}
We study the possibility of spatially non-uniform ground state in
 (1+1)-dimensional models with quartic fermi interactions at finite
 fermion densities by introducing chemical potential $\mu$.
We examine the chiral Gross-Neveu model and the Cooper pair model as
 toy models of the chiral symmetry breaking and the difermion pair
 condensates which are presumed to exist in QCD.
We confirm in the chiral Gross-Neveu model that the ground state has
 a crystalline structure in which the chiral condensate oscillates in
 space with wave number $2\mu$.
Whereas in the Cooper pair model we find that the vacuum structure is
 spatially uniform.
Some discussions are given to explain this difference.
\end{abstract}


%%%%%%%%%%^%%%%%%%%%^%%%%%%%%%^%%%%%%%%%^%%%%%%%%%^%%%%%%%%% Section 1
\section{Introduction}
\label{sec:intro}

Illuminating the vacuum structure of QCD at high baryonic densities,
 which should be useful to understand physics of compact stars and heavy
 ion collisions, remains challenge for particle and nuclear theorists.
One of the most interesting in the related subjects is the color
 superconductivity of quarks \cite{CS1,CS2,CS3}.
Quark Cooper pairs condense due to the BCS instability of Fermi surface
 at high densities and low temperatures, while chiral condensates is
 dynamically generated by particle-antiparticle pairing at low
 temperatures and densities.
Since quarks have a color component of the $SU(3)$ gauge group, the
 ground state of QCD is a superconductor with spontaneous breaking of
 color symmetry as a result of diquark condensates.

In this paper, we address the question of spatial variations of
 fermionic condensates, the problem first investigated by Deryagin,
 Grigoriev and Rubakov (DGR) in QCD with infinite number of colors
 $N_c$ \cite{DGR}.
Working in the perturbative regime $g^2N_c\ll 1$ ($g$ is the gauge
 coupling) and by using the variational method, DGR discovered a
 previously unsuspected (in the context of QCD) instability of the
 Fermi surface. Namely, they found that an instability exists in the
 large $N_c$ QCD at finite densities which triggers condensate of
 particle-hole pairs (Overhauser effect) having almost the same momentum
 $\Vec p$ with modulus $|\Vec p|=\mu$, where $\mu$ is the chemical
 potential.
They showed that the ground state of the theory is spatially nonuniform,
 and has a periodically varying chiral condensate with wave number
 $2\mu$.

Following DGR, analyses which utilize either renormalization group
 analysis \cite{SS} or Wilsonian effective action formalism \cite{PRWZ}
 as well as the Nambu-Gorkov formalism \cite {RSZ} were attempted to
 examine which effect, the Overhauser or the BCS effects wins at large,
 at finite $N_c$.
It was shown in these analyses that the BCS effect overtakes the
 Overhauser effect in weak coupling regime, while Overhauser wins
 against BCS in strong coupling regime.
More recently, the possibility of crystalline ground state was
 investigated in the context of color superconductivity
 \cite{ABR,BKRS,LRS}.
The underlying mechanism which leads to the crystalline structure is a
 mismatch in the fermi surface of up and down quarks, and is apparently
 quite different from one for the periodic chiral condensate discussed
 by DGR.

To have better understanding of the relationship between these related
 but different phenomena, it is desirable to have a theoretical
 laboratory to investigate the underlying mechanism of these different
 phenomena.
In this paper, we examine two simple (1+1)-dimensional models with
 quartic fermi interactions which would serve for this purpose.
We examine the chiral Gross-Neveu model \cite{GN} and the Cooper pair
 model proposed by Chodos, Minakata and Cooper \cite{CMC}.

The chiral Gross-Neveu model \cite{GN} is a natural choice as the
 two-dimensional model which displays dynamical breaking of continuous
 chiral symmetry.
It is also well known that the Gross-Neveu model at infinite number of
 flavor has a number of similar properties with QCD; the
 renormalizability, the asymptotic freedom, and the similar phase
 structure on $\mu$-$T$ plane with those suspected for the two-flavor
 QCD.

The Cooper pair model \cite{CMC} is a toy model for "color
 superconductivity" in QCD but without color degrees of freedom.
It nevertheless serves as a renormalizable asymptotically free model
 field theory which admits Lorentz scalar fermion pair condensate.
The model possesses $O(N)$ flavor symmetry and $U(1)$ symmetry of
 fermion number, and the authors of Ref.~\cite{CMC} have shown that at
 the large-$N$ limit the $U(1)$ symmetry is dynamically broken by the
 Lorentz singlet di-fermion condensate.
It was also shown that the model, when coupled with the Gross-Neveu
 model, has a very similar phase structure with that of QCD with two
 flavor of quarks \cite{CCMMS}.

Strictly speaking, continuous symmetry such as the $U(1)$ symmetries of
 chirality or fermion number cannot be violated by the
 Coleman-Mermin-Wagner theorem \cite{CMW} in (1+1) dimensions.
However, it is well understood by now in what sense the theorem is
 "invalidated" in the large-$N$ limit.
As Witten argued \cite{witten}, the correlation functions of order
 parameters have power law behavior at large distances, and the power is
 inversely proportional to $N$.
Therefore, if we take the limit $N \rightarrow \infty$ first and then
 send spatial distance between two order parameters infinity to use the
 cluster property, we formally obtain nonvanishing vacuum expectation
 value of the order parameter.
The procedure seems to give a consistent field theory, which is usuable
 as a theoretical laboratory for investigating various properties of
 condensate phenomena.

We develop a systematic method for investigating the possibilities of
 spatial varing condensates, and then apply the formalism to discuss the
 chiral Gross-Neveu model and the Cooper pair model in a unified way.
In a recent paper, Sch\"{o}n and Thies \cite{ST} treated the chiral
 Gross-Neveu model and concluded that the model has chiral wave
 condensates with a wave number $2\mu$.
Our formalism differs from theirs on several respects, in particular on
 how to regulate ultraviolet divergence.
We will reexamine the chiral Gross-Neveu model in the light of our
 formalism and confirm their results.
On the other hand, We obtain a spatially uniform fermion pair condensate
 as the stable ground state of the Cooper pair model, in agreement with
 \cite{CMC}.

In Sec.~\ref{sec:hamiltonian} we introduce the quasi-particle picture
 and the Hamiltonian is diagonalized in terms of the quasi-particles.
In Sec.~\ref{sec:GN} the effective potential is derived by taking the
 expectation values of the Hamiltonian with respect to properly defined
 variational vacua and we search for a vacuum solution.
In Sec.~\ref{sec:BV} we relate the quasi-particles to the free particles
 through the Bogoliubov-Valatin transformation, which leads us to the
 fact that the spatial variations of the chiral condensate yield due to
 the nonvanishing momentum of particle-hole/antiparticle pairs.
We also discuss validity of the ultraviolet regularization by
 considering the fermion number of vacuum.
In Sec.~\ref{sec:CP} the similar discussion to the chiral Gross-Neveu
 model is applied to the Cooper pair model.
Sec.~\ref{sec:conc} is devoted to conclusions.
In Appendix we discuss the ultraviolet regularization scheme in detail.

%%%%%%%%%%^%%%%%%%%%^%%%%%%%%%^%%%%%%%%%^%%%%%%%%%^%%%%%%%%% Section 2
\section{The Hamiltonian of chiral Gross-Neveu model}
\label{sec:hamiltonian}

First of all, we will focus our attention on the chiral Gross-Neveu
 model in which the Lagrangian is given by
%%%%%%%%%%%%%%%%%
\begin{equation}%
  {\cal L} = \psb\left(i\ds+\gamma^0\mu\right)\ps + \frac12g^2
     \abs{\psb(1+\gamma^5)\ps}^2.
\label{GN-lag}
\end{equation}%
%%%%%%%%%%%%%%%
Here, $\ps$ is a two-component spinor with a flavor index of $SU(N)$.
Repeated flavor indices in Eq.~(\ref{GN-lag}) are meant to be summed.
To solve the model, $N$ is sent to an infinite, keeping $\lambda_0=g^2N$
 fixed.
We also introduce a chemical potential $\mu$ which is a constant source
 for fermion number in order to take account of the finite fermion
 density of the system.
The Lagrangian has a $U(1)$ chiral symmetry, which is dynamically broken
 at zero temperature and $\mu=0$ \cite{GN}.
Under the assumption of spacially uniform ground state the symmetry is
 shown to be restored at finite temperature and/or chemical potential
 \cite{Wo}.
A recent work that relaxes the assumption of spatial uniformity to
 accomodate a periodically varing one\cite{ST} has shown that a chiral
 crystalline state with
 $\langle\psb(1+\gamma^5)\ps\rangle\propto e^{2i\mu x}$ has lower energy
 than uniform chiral symmetry breaking or unbroken state.
Using this model, we will first develop a method for investigating
 possibilities of spatially varing condensates, which can also be
 applied to the Cooper pair model.

Our strategy is to use the auxiliary field method.
Following standard techniques \cite{GN} we can add the term involving
 complex, auxiliary field $\sigma(x)$ without affecting on the dynamics
 of the theory:
%%%%%%%%%%%%%%%%%
\begin{equation}%
  {\cal L} \rightarrow
    {\cal L}-\frac1{2g^2}\abs{\sigma+g^2\psb(1+\gamma^5)\ps}^2.
\label{axpot}
\end{equation}%
%%%%%%%%%%%%%%%
An effective action of the model is obtained by doing the functional
 integral with respect to both auxiliary and fermion fields.
In the large-$N$ limit, however, $\sigma(x)$ can be regarded as a
 classical, background field for fermions because it is determined by
 the method of stationary phase.
Instead of integrating out the fermion fields in the effective action,
 we diagonalize the fermionic Hamiltonian in the presence of the
 background field $\sigma(x)$ in terms of the creation and annihilation
 operators of quasi-particles.
The free energy is then obtained as expectation values of the
 Hamiltonian with respect to variational ground states for each
 $\sigma$.
At this moment, since a $U(1)$ chiral transformation including the
 auxiliary field is indicated by $\ps\rightarrow e^{i\alpha\gamma^5}\ps$
 and $\sigma\rightarrow e^{2i\alpha}\sigma$ ($\alpha$ is an arbitrary
 constant), non-vanishing classical field $\sigma$ implies spontaneous
 symmetry breaking of the $U(1)$ chiral symmetry.

To analyze the chiral Gross-Neveu model and the Cooper pair model in
 parallel, it is convenient to introduce the charge conjugation field
 defined by
%%%
\[%
  \psch(x)=C\psb^T(x),
\]%
%%%
 where $C$ is a $2\times2$ matrix with the properties (see, for example,
 Bjorken and Drell \cite{BD})
%%%
\[%
  C^{-1}\gamma^\mu C=-\gamma^{\mu T},\;\;\;C=-C^T.
\]%
%%%
In our convention of Dirac matrices, $\gamma^0=\sigma_1$,
 $\gamma^1=-i\sigma_2$ and $\gamma^5=\sigma_3$ ($\sigma$'s are Pauli
 matrices), $C=-\gamma^1$ and charge conjugation field is described as
%%%%%%%%%%%%%%%%%
\begin{equation}%
  \psch(x)=\gamma^5\ps^\ast(x).
\label{cc}
\end{equation}%
%%%%%%%%%%%%%%%
By using the integral identity
%%%%%%%%%%%%%%%%%
\begin{equation}%
   \frac12\int d^2x\psb \left(i\ds+\gamma^0\mu\right)\ps
  =\frac12\int d^2x\pscb\left(i\ds-\gamma^0\mu\right)\psch,
\label{kinetic}
\end{equation}%
%%%%%%%%%%%%%%%
 the fermionic Hamiltonian density corresponding to the Lagrangian of
 the chiral Gross-Neveu model can be written as
%%%%%%%%%%%%%%%%%
\begin{eqnarray}%
 {\cal H} &=&\frac{\partial{\cal L}}{\partial\dot{\psi_i}}\psi_i
   + \frac{\partial{\cal L}}{\partial\dot{\psi_i^c}}\psi_i^c
   - {\cal L}                                              \nonumber \\
 &=& \frac{\asigma^2}{2g^2}
 + \frac12
       \left[\psd,\pscd\right]
       \left[\begin{array}{cc}
         h&0\\0&h^c\\
       \end{array}\right]
       \left[\begin{array}{c}
  \ps \\ \psch
       \end{array}\right],
\label{hamilt}
\end{eqnarray}%
%%%%%%%%%%%%%%%
 where the first quantized Hamiltonian
 ${\hat H}=\left[\begin{array}{cc} h&0\\0&h^c\\\end{array}\right]$
 is given by
%%%%%%%%%%%%%%%%%
\begin{eqnarray}%
 h   &=& -i\gamma^5\partial_x-\mu+\gamma^0\mbox{Re}\;\sigma
           -i\gamma^1\mbox{Im}\;\sigma                      \nonumber \\
 h^c &=& -i\gamma^5\partial_x+\mu+\gamma^0\mbox{Re}\;\sigma
           +i\gamma^1\mbox{Im}\;\sigma
\end{eqnarray}%
%%%%%%%%%%%%%%%
Note that the sign of $\mbox{Im}\,\sigma$ term flips because the charge
 conjugation is a kind of chiral conjugation (see, Eq.~(\ref{cc})) and
 the imaginary part of an auxiliary field $\sigma$ transforms as
 $\psb\gamma^5\ps$.

Our next task is to identify the eigenstates of Eq.~(\ref{hamilt}) and to
 expand the fermion field in terms of them.
To do this, we assume that $\sigma(x)$ has the following static standing
 wave form:
%%%%%%%%%%%%%%%%%
\begin{equation}%
  \sigma(x)=\asigma e^{2iKx},
\label{sigma-ans}
\end{equation}%
%%%%%%%%%%%%%%%
 where $\asigma$ and $K$ are arbitary constants which will be determined
 later.
We solve eigenvalue problem $\hat{H}\Psi=\omega_n\Psi$ by Fourier
 expanding with the chiral phase factor $e^{\pm iKx\gamma^5}$.
We obtain a complete, orthonormal set of eigen functions
 $\Psi=\left[\psi_n(x),\psi_n^c(x)\right]^T$.
It consists of four types of continuum state labeled as
 $n=(p,a),(p,b),(-p,\bar{a}),(-p,\bar{b})$ with the following eigenvalus
 ($p$ is the Fourier mode and we call this type of eigenstates
 ``branch'', e.g., $a$-branch):
%%%%%%%%%%%%%%%%%
\begin{equation}%
\begin{array}{lclcl}
  \omega_{pa}=\epsilon_p-\mu_\ast &:&
    \psi_{pa}(x)  =e^{i(p+\gamma^5K)x}u_p &,&
    \psi_{pa}^c(x)=0, \\
  \omega_{pb}=\epsilon_p+\mu_\ast &:&
    \psi_{pb}(x)  =0 &,&
    \psi_{pb}^c(x)=e^{i(p-\gamma^5K)x}u_p, \\
  \omega_{-p\bar{a}}=-\epsilon_p+\mu_\ast &:&
    \psi_{-p\bar{a}}(x)  =0 &,&
    \psi_{-p\bar{a}}^c(x)=e^{i(p-\gamma^5K)x}u_{-p}^c, \\
  \omega_{-p\bar{b}}=-\epsilon_p-\mu_\ast &:&
    \psi_{-p\bar{b}}(x)  =e^{i(p+\gamma^5K)x}u_{-p}^c &,&
    \psi_{-p\bar{b}}^c(x)=0. \\
\end{array}
\label{eigst}
\end{equation}%
%%%%%%%%%%%%%%%
Here, $\epsilon_p=\sqrt{p^2+\asigma^2}$ and $\mu_\ast=\mu-K$.
The two component spinors in Eqs.~(\ref{eigst}) are
%%%%%%%%%%%%%%%%%
\begin{eqnarray}%
  u_p &=& \left[\begin{array}{c} \cos\frac12\theta_p \\
            \sin\frac12\theta_p \\ \end{array}\right]
    = \frac1{\sqrt{2\epsilon_p}}\left[\begin{array}{c}
          \sqrt{\epsilon_p+p} \\ \sqrt{\epsilon_p-p} \\
        \end{array}\right],                               \nonumber \\
  u_{-p}^c = \gamma^5u_{-p}^\ast
   &=& \left[\begin{array}{c} \sin\frac12\theta_p \\
           -\cos\frac12\theta_p \\ \end{array}\right]
    =  \frac1{\sqrt{2\epsilon_p}}\left[\begin{array}{c}
           \sqrt{\epsilon_p-p} \\ -\sqrt{\epsilon_p+p} \\
         \end{array}\right].
\label{eig-spi}
\end{eqnarray}%
%%%%%%%%%%%%%%%
Note that the wave functions are no longer momentum eigenstates in the
 presense of $K$ because of the chiral phase factor.
While $\omega$'s may be regarded as the energies of a fermion or an
 antifermion with momentum $p$, mass $\asigma$ and chemical potential
 $\mu_\ast$, they must be interpreted merely as quasi-particle
 dispersion relations (especially, $p$ is not a momentum of particles).

The fermion field can be expanded in terms of eigenstates (\ref{eigst}).
The coefficients of four branches of eigenstates are not independent
 with each other, but $a(b)$- and $\bar{a}(\bar{b})$-branches must be
 hermitian conjugate with each other to assure that
 $\psch=\gamma^5\ps^\ast$.
The second quantized fermion field is then given by
%%%%%%%%%%%%%%%%%
\begin{equation}%
  \left[\begin{array}{c} \ps(x) \\ \psch(x) \end{array}\right] =
  \int\frac{dp}{2\pi}\left[\begin{array}{c}
    e^{i(p+\gamma^5K)x}\Bigl(u_p\aap+u_{-p}^c\bdm\Bigr) \\
    e^{i(p-\gamma^5K)x}\Bigl(u_{-p}^c\adm+u_p\bbp\Bigr) \\
  \end{array}\right].
\label{mode-ex}
\end{equation}%
%%%%%%%%%%%%%%%
Here, $\aap$, $\bbp$ and their hermitian conjugates are creation and
 annihilation operators of quasi-particles.
To hold the canonical anticommutation relations,
 $\{\psi_i(x),\psi_j^{\dag}(y)\}=\delta_{ij}\delta(x-y)$, they must obey
 anticommutation relations;
 $\{a_{pi},a_{qj}^{\dag}\}=\{b_{pi},b_{qj}^{\dag}\}
   =2\pi\delta(p-q)\delta_{ij}$,
 and all other anticommutators vanish.

Using the decomposition into Fourier space, the field operators can be
 expressed in terms of quasi-particle operators, $a$'s and $b$'s.
For example, Hamiltonian and fermion number operator become
%%%%%%%%%%%%%%%%%
\begin{eqnarray}%
  H &=& \int dx{\cal H}                                        \nonumber \\
    &=& NL\frac{\asigma^2}{2\lambda_0}+\frac12\int\frac{dp}{2\pi}\left(
          \omega_{ p     a }\adp\aap+\omega_{ p     b }\bdp\bbp
         +\omega_{-p\bar{a}}\aam\adm+\omega_{-p\bar{b}}\bbm\bdm
        \right),                                        \label{ch-hamil} \\
  J^0 &=& \frac12\int dx\left[\psd\ps-\pscd\psch\right]        \nonumber \\
      &=& \frac12\int\frac{dp}{2\pi}\left(
            \adp\aap-\bdp\bbp-\aam\adm+\bbm\bdm \right),
\label{charge}
\end{eqnarray}%
%%%%%%%%%%%%%%%
 where $L=\int{dx}$ is the spatial volume.
It is found that $\adp$ is the annihilation operator of the
 $\bar a$-branch simultaneously with the creation operator of the
 $a$-branch.
It should be also noticed that the quasi-particle operators diagonalize
 the fermion number operator differently from the Cooper pair model.
This becomes the crucial point to distinguish between the two models, as
 one can find below.

%%%%%%%%%%^%%%%%%%%%^%%%%%%%%%^%%%%%%%%%^%%%%%%%%%^%%%%%%%%% section 3
\section{Effective potential and Vacuum structure for chiral Gross-Neveu
 model}
\label{sec:GN}

In this section we want to construct and analyze the effective potential
 for the chiral Gross-Neveu model.
We begin by calculating expectation values of the Hamiltonian with
 respect to variational ground state which is defined by filling the
 negative energy states;
%%%%%%%%%%%%%%%%%
\begin{equation}%
\begin{array}{ccccccc}
  \omega_{pa}=-\omega_{p\bar a}>0 &:& \aap\vacs=0 &,&
  \omega_{pb}=-\omega_{p\bar b}>0 &:& \bbp\vacs=0 \\
  \omega_{pa}=-\omega_{p\bar a}<0 &:& \adp\vacs=0 &,&
  \omega_{pb}=-\omega_{p\bar b}<0 &:& \bdp\vacs=0 \\
\end{array}.
\label{sig-vac}
\end{equation}%
%%%%%%%%%%%%%%%
Evaluating the expectation value of the Hamiltonian with respect to
 $\vacs$, the effective potential becomes (for brevity, we denote
 $\langle{\cal O}\rangle=\langle\sigma|{\cal O}\vacs$)
%%%%%%%%%%%%%%%%%
\begin{eqnarray}%
  {\cal V}(\asigma,K) &=& \frac{\langle H\rangle}{NL} \nonumber \\
    &=& \frac{\asigma^2}{2\lambda_0}
       +\frac12\int\frac{dp}{2\pi}\omega_{-p\bar{a}}
       +\frac12\int\frac{dp}{2\pi}\omega_{-p\bar{b}}
       +\int\limits_{\omega_{pa}<0}\frac{dp}{2\pi}\omega_{pa}
       +\int\limits_{\omega_{pb}<0}\frac{dp}{2\pi}\omega_{pb},
\label{effec}
\end{eqnarray}%
%%%%%%%%%%%%%%%
 where the region of integration in the second and the third terms in
 the last line consists of whole range of $p$.
The last two integrals come from ``no-pairing'' regions in which the
 modes does not give any contribution to dynamical generation of chiral
 condensate (see, e.g, Eq.~(\ref{chi-cond})).
Under the condition for absence of no-paring regions,
 $\mu_\ast^2<\asigma^2$, there is no effect from free fermions and these
 integrals vanish.

Eq.~(\ref{effec}) contains several divergences at ultraviolet regions;
 for example, if we introduce the naive Fourier mode cutoff (as we took
 care before, $p$ is not a momentum of quasi-particle),
 $\abs{p}<\Lambda/2$, then the divergent terms come from first two
 integrals in Eq.~(\ref{effec}):
%%%%%%%%%%%%%%%%%
\begin{eqnarray}%
  \int\limits_{-\Lambda/2}^{\Lambda/2}\frac{dp}{2\pi}\omega_{-p\bar{a}}
    &=& -\frac{\Lambda^2}{8\pi}+\frac{\mu_\ast}{2\pi}\Lambda
        -\frac{\asigma^2}{4\pi}\ln\frac{\Lambda^2}{\asigma^2}
        -\frac{\asigma^2}{2\pi}+{\cal O}(1/\Lambda)           \nonumber \\
  \int\limits_{-\Lambda/2}^{\Lambda/2}\frac{dp}{2\pi}\omega_{-p\bar{b}}
    &=& -\frac{\Lambda^2}{8\pi}-\frac{\mu_\ast}{2\pi}\Lambda
        -\frac{\asigma^2}{4\pi}\ln\frac{\Lambda^2}{\asigma^2}
        -\frac{\asigma^2}{2\pi}+{\cal O}(1/\Lambda)
\label{div-part0}
\end{eqnarray}%
%%%%%%%%%%%%%%%
The quadratic and linear divergences depend on the cutoff scheme,
 although the logarithmic divergence does not.
These divergences therefore lead to nontrivial issue for the ultraviolet
 regularization.
We will show at the next section that the naive Fourier mode cutoff has
 problems with the number of fermions floating in the vacuum.
We employ a different cutoff scheme which is free from the trouble.
As will be explained somewhat in detail in Appendix, it involves the
 branch dependent cutoff prescription:
%%%%%%%%%%%%%%%%%
\begin{equation}%
  \mbox{$a(\bar{a})$-branch}:
    \left[-\frac{\Lambda}{2}+K,\frac{\Lambda}{2}-K\right],\;\;\;
  \mbox{$b(\bar{b})$-branch}:
    \left[-\frac{\Lambda}{2}-K,\frac{\Lambda}{2}+K\right].
\label{uv-cut}
\end{equation}%
%%%%%%%%%%%%%%%
The reader may wonder why one should take such branch-dependent shift on
 the ultraviolet cutoff.
In fact, it is shown in Refs.~\cite{RN} that the similar shift resulting
 from a phase shift in the normal modes of fluctuations around the
 soliton gives correct quantum corrections to the kink mass in the
 two-dimentional $\phi^4$-theory.

By using the cutoff prescription (\ref{uv-cut}), first two integrals in
 the effective potential (\ref{effec}) become
%%%%%%%%%%%%%%%%%
\begin{eqnarray}%
  \int\limits_{-\Lambda/2+K}^{\Lambda/2-K}\frac{dp}{2\pi}\omega_{-p\bar{a}}
    &=& -\frac{\Lambda^2}{8\pi}+\frac{\mu}{2\pi}\Lambda
        -\frac{\asigma^2}{4\pi}\ln\frac{\Lambda^2}{\asigma^2}
        -\frac{\asigma^2}{2\pi}+\frac{\mu_\ast^2}{2\pi}
        -\frac{\mu^2}{2\pi}+{\cal O}(1/\Lambda)                  \nonumber
\\
  \int\limits_{-\Lambda/2-K}^{\Lambda/2+K}\frac{dp}{2\pi}\omega_{-p\bar{b}}
    &=& -\frac{\Lambda^2}{8\pi}-\frac{\mu}{2\pi}\Lambda
        -\frac{\asigma^2}{4\pi}\ln\frac{\Lambda^2}{\asigma^2}
        -\frac{\asigma^2}{2\pi}+\frac{\mu_\ast^2}{2\pi}
        -\frac{\mu^2}{2\pi}+{\cal O}(1/\Lambda)
\label{div-part}
\end{eqnarray}%
%%%%%%%%%%%%%%%
One notices by comparing Eqs.~(\ref{div-part}) to (\ref{div-part0}) that
 additional finite contributions arise from the lineary or quadraticaly
 divergent terms due to the shifts in cutoff.
The quadratic divergence in Eq.~(\ref{div-part}) is an uninteresting
 constant which can be subtracted off and the linear divergences are
 exactly cancelled between two integrals.
The remaining logarithmic divergence in Eqs.~(\ref{div-part}) can be
 removed, as usual, by the renormalization of the coupling constant.
Since they contain only $\asigma$ dependence, it is enough to carry it
 out in the case of $K=\mu=0$.
Therefore, we renormalize by demanding that the renormalized coupling
 constant $\lambda$ satisfies
%%%%%%%%%%%%%%%%%
\begin{equation}%
  \frac{\partial^2{\cal V}}{\partial\asigma^2}\Bigg|_{
    {\asigma=m_0}\atop{K=\mu=0}
     }=\frac1\lambda.
\label{ren-cond}
\end{equation}%
%%%%%%%%%%%%%%%
Here $\asigma=m_0$ designates an arbitary renormalization point on which
 the coupling constant depends.
Using this conditions to solve for $\lambda_0$ in terms of $\lambda$
 yields the renormalized form of the effective potential:
%%%%%%%%%%%%%%%%%
\begin{eqnarray}%
  \mu_\ast^2<\asigma^2 &:& {\cal V}=\frac{\asigma^2}{4\pi}\left[
     \frac{2\pi}{\lambda}-3+\ln\frac{\asigma^2}{m_0^2}\right]
    +\frac{\mu_\ast^2}{2\pi}-\frac{\mu^2}{2\pi},              \nonumber \\
  \mu_\ast^2>\asigma^2 &:& {\cal V}=\frac{\asigma^2}{4\pi}\left\{
     \frac{2\pi}{\lambda}-3+\ln\frac{\mu_\ast^2}{m_0^2}\left[
       1+\sqrt{1-\frac{\asigma^2}{\mu_\ast^2}}\right]^2\right\}
                                                        \label{R-poten} \\
 && \hspace{144pt}
    +\frac{\mu_\ast^2}{2\pi}\left[1-\sqrt{1-
       \frac{\asigma^2}{\mu_\ast^2}}\right]-\frac{\mu^2}{2\pi}. \nonumber
\end{eqnarray}%
%%%%%%%%%%%%%%%
When $K=0$, Eqs.~(\ref{R-poten}) turns into the well-known effective
 potential assuming the homogeneity of chiral condensates \cite{Wo}.
If we set $K=\mu$, the effective potential is reduced to one in
 Ref.~\cite{ST}.

\begin{figure}
\begin{center}
 \epsfig{file=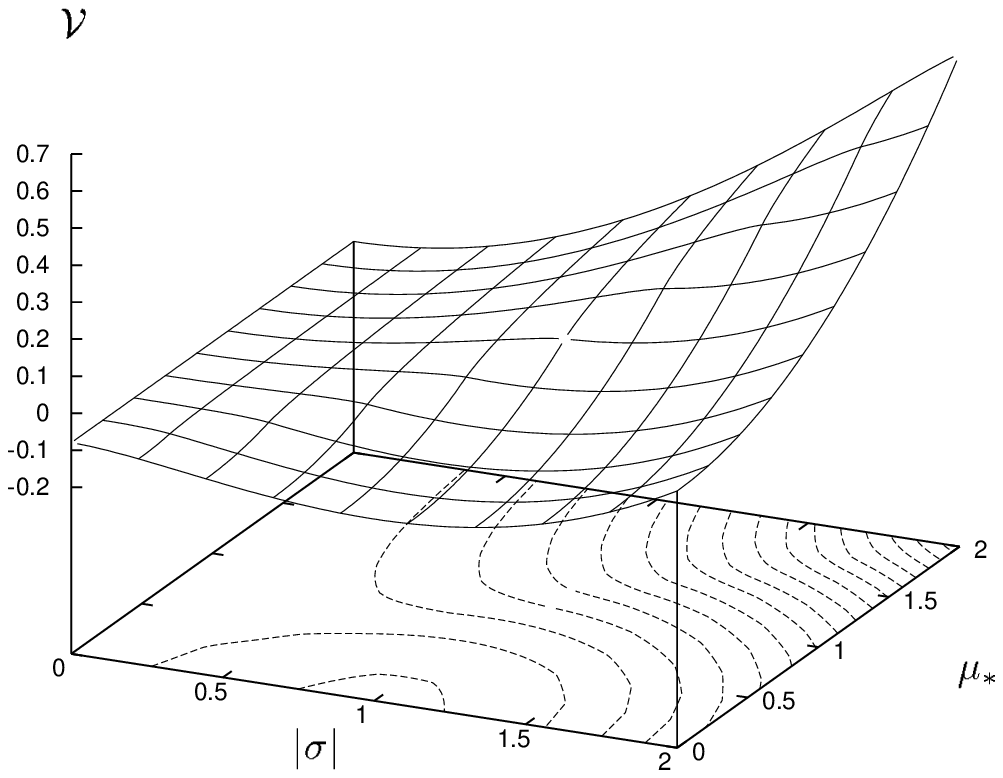}
\end{center}
\caption{The effective effective potemtial for the chiral Gross-Neveu
 model.
 Each axis is scaled by $m=m_0e^{1-\pi/\lambda}$.
 The contour of the effective potential is depicted on the bottom of 3D
 plot.
 For the Cooper pair model, replace $\asigma$ and $\mu_\ast$ by
 $\adelta$ and $K$, respectively.}
\label{fig:potential}
\end{figure}

It is easy to show that $\asigma=m_0e^{1-\pi/\lambda}$,
 $\mu_\ast=\mu-K=0$ is the solution of the gap equations derived from
 the effective potential, $\frac{\partial{\cal V}}{\partial\asigma}
    =\frac{\partial{\cal V}}{\partial K}=0$.
It give the minmum value of effective potential as one can see in
 Fig.~\ref{fig:potential}.
Therefore, the vacuum structure is characterized by
%%%%%%%%%%%%%%%%%
\begin{equation}%
  \sigma(x)=\left(m_0e^{1-\pi/\lambda}\right)e^{2i\mu x}.
\label{vacsig}
\end{equation}%
%%%%%%%%%%%%%%%
Using the field decomposition (\ref{mode-ex}), one can evaluate the
 chiral condensate of vacuum explicitly,
%%%%%%%%%%%%%%%%%
\begin{eqnarray}%
  \langle\psb(1+\gamma^5)\ps\rangle
   &=&-N\asigma e^{2iKx}\frac1{2\pi}\Biggl[\int\frac{dp}{\epsilon_p}
        -\int\limits_{\omega_{pa}<0}\frac{dp}{\epsilon_p}
        -\int\limits_{\omega_{pb}<0}\frac{dp}{\epsilon_p}\Biggr] \nonumber
\\
   &=&-N\sigma(x)/\lambda_0,
\label{chi-cond}
\end{eqnarray}%
%%%%%%%%%%%%%%%
 where the vacuum value of $\sigma(x)$, Eq.~(\ref{vacsig}), is used in
 the second line of the equation.
The equation guarantees to hold the equation of motion for $\sigma(x)$
 derived from the Lagrangian (\ref{axpot}).
Consequently, we have reproduced the result in Ref.~\cite{ST} which the
 phase standing chiral density wave with a wave number $2\mu$ is stable
 as compared to uniformally chiral symmetry broken phase or the
 symmetric phase.
Moreover, we have demonstrated that the condensate of the wave number
 $2\mu$ arises as the unique minimum of the effective potential.
We present arguments to clarify the reasons why the wave number is
 $2\mu$ at the next section.

%%%%%%%%%%^%%%%%%%%%^%%%%%%%%%^%%%%%%%%%^%%%%%%%%%^%%%%%%%%% Section 4
\section{Relation to the Overhauser effect}
\label{sec:BV}

\begin{figure}
\begin{center}
 \epsfig{file=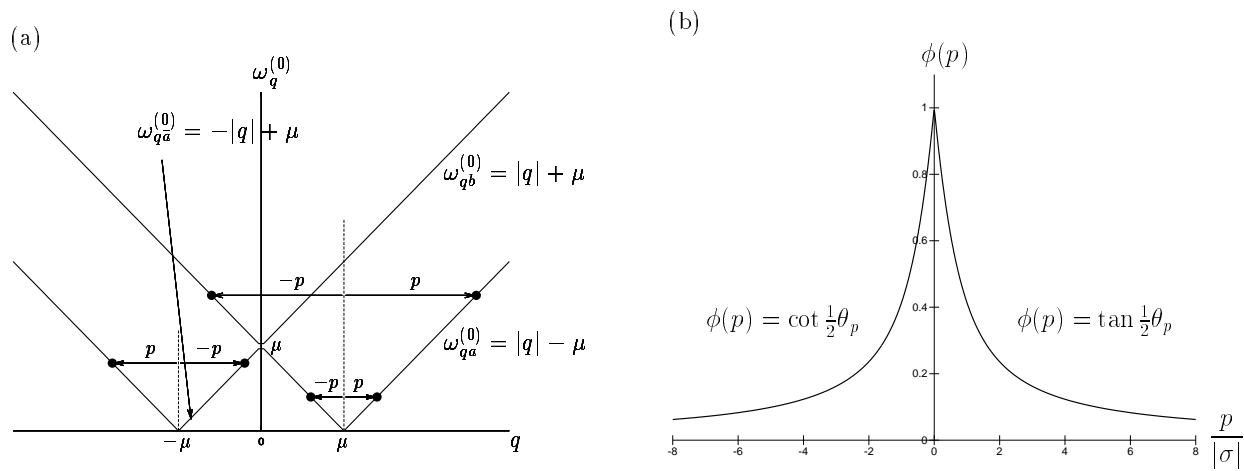,width=\linewidth}
\end{center}
\caption{(a) The manner of pairing for the chiral Gross-Neveu vacuum.
 $\omega_q^{(0)}$ is free particle energy; thus, $\omega_{qa}^{(0)}$ is
 of the fermion, $\omega_{q\bar a}^{(0)}$ is of the hole and
 $\omega_{qb}^{(0)}$ is of the antifermion.
 The pairing with total momentum $2\mu$ ($p>0$) or $-2\mu$ ($p<0$) yields. 
 (b) Pairing wave function $\phi(p)$.}
\label{fig:pairing}
\end{figure}

To understand why the chiral condensate oscillates in space with wave
 number $2\mu$, we start by considering that the field expansion,
 Eq.~(\ref{mode-ex}), has a connection with an interaction-free
 description through the Bogoliubov-Valatin transformation.
For brevity, we drop the flavor indices and set $L$ to unity.
Setting $\asigma=K=0$ in Eq.~(\ref{mode-ex}), the free field expansion
 is
%%%%%%%%%%%%%%%%%
\begin{equation}%
  \psi(x)=\int\frac{dq}{2\pi}e^{iqx}\left[\begin{array}{c}
    a_{q}^{(0)}\Theta( q)+b_{-q}^{(0)\dag}\Theta(-q) \\
    a_{q}^{(0)}\Theta(-q)-b_{-q}^{(0)\dag}\Theta( q) \\ \end{array}\right],
\label{free-ex}
\end{equation}%
%%%%%%%%%%%%%%%
 where the subscript $(0)$ implies the operators corresponding to free
 massless fermion and antifermion, and $\Theta$ denotes the Heaviside
 step function.
By equating Eq.~(\ref{free-ex}) to Eq.~(\ref{mode-ex}), the
 Bogoliubov-Valatin transformation reads
%%%%%%%%%%%%%%%%%
\begin{eqnarray}%
   K<p    &:& \left\{\begin{array}{l}
                a_p = a_{ p+K}^{(0)    }\cos\frac12\theta_p
                     -b_{-p+K}^{(0)\dag}\sin\frac12\theta_p,\\
                b_p = b_{ p-K}^{(0)    }\cos\frac12\theta_p
                     -a_{-p-K}^{(0)\dag}\sin\frac12\theta_p,\\
              \end{array}\right. \nonumber        \\
  -K<p< K &:& \left\{\begin{array}{l}
                a_p = a_{ p+K}^{(0)    }\cos\frac12\theta_p
                     +a_{ p-K}^{(0)    }\sin\frac12\theta_p,\\
                b_p = a_{-p+K}^{(0)\dag}\cos\frac12\theta_p
                     -a_{-p-K}^{(0)\dag}\sin\frac12\theta_p,\\
              \end{array}\right. \label{BV-trans} \\
     p<-K &:& \left\{\begin{array}{l}
                a_p = b_{-p-K}^{(0)\dag}\cos\frac12\theta_p
                     +a_{ p-K}^{(0)    }\sin\frac12\theta_p,\\
                b_p = a_{-p+K}^{(0)\dag}\cos\frac12\theta_p
                     +b_{ p+K}^{(0)    }\sin\frac12\theta_p.\\
              \end{array}\right. \nonumber
\end{eqnarray}%
%%%%%%%%%%%%%%%
We can then relate the variational ground state $\vacs$ to the normal
 state $\vac$.
The latter is annihilated by $a^{(0)}\:(a^{(0)\dag})$ and $b^{(0)}$
 above (below) the Fermi surface:
%%%%%%%%%%%%%%%%%
\begin{equation}%
\begin{array}{ccl}
  \abs{q}>\mu &:& a_q^{(0)    }\vac=b_q^{(0)}\vac=0, \\
  \abs{q}<\mu &:& a_q^{(0)\dag}\vac=b_q^{(0)}\vac=0, \\
\end{array}
\label{nor-vac}
\end{equation}%
%%%%%%%%%%%%%%%
 (We implicitly assume that $\mu>0$ such that the fermionic Fermi
 surface is formed).
For the case of $K=\mu$ in particular, one is able to show that $\vacs$
 can be written in a form of the BCS state,
%%%%%%%%%%%%%%%%%
\begin{equation}%
  \vacs=\prod_{  p<-\mu}A_p^{-\dag}\prod_{-\mu<p<0}B_p^{-\dag}
        \prod_{0<p< \mu}B_p^{+\dag}\prod_{ \mu<p  }A_p^{+\dag}\vac,
\label{BV-vac}
\end{equation}%
%%%%%%%%%%%%%%%
 where
%%%%%%%%%%%%%%%%%
\begin{equation}%
\begin{array}{l}
  A_p^{+\dag}=\cos\frac12\theta_p+\sin\frac12\theta_p
                 a_{p+\mu}^{(0)\dag}b_{-p+\mu}^{(0)\dag}, \\
  B_p^{+\dag}=\cos\frac12\theta_p+\sin\frac12\theta_p
                 a_{p-\mu}^{(0)}a_{p+\mu}^{(0)\dag},      \\
  B_p^{-\dag}=\sin\frac12\theta_p+\cos\frac12\theta_p
                 a_{p+\mu}^{(0)}a_{p-\mu}^{(0)\dag}.      \\
  A_p^{-\dag}=\sin\frac12\theta_p+\cos\frac12\theta_p
                 b_{-p-\mu}^{(0)\dag}a_{p-\mu}^{(0)\dag}, \\
\end{array}
\end{equation}%
%%%%%%%%%%%%%%%
It can be easily verified that $\vacs$ in Eq.~(\ref{BV-vac}) satisfies
 the definition of it, Eq.~(\ref{sig-vac}), by using
 Eqs.~(\ref{BV-trans}) and (\ref{nor-vac}).
In the region of $p>\mu$, $\vacs$ comprises with the
 particle-antiparticle pairing as in the usual chiral condensate vacua
 but with pair momentum $2\mu$  in our case.
On the other hand, in the region of $0<p<\mu$, $\vacs$ comprises the
 particle-hole pairing with the momentum $(p+\mu)$ and $-(p-\mu)$,
 respectively.
The manner of pairing is depicted in Fig.~\ref{fig:pairing}.
This figure tells us that either the particle-hole or the
 particle-antiparticle pairing across the Fermi surface replaces
 particle-antiparticle pairing across the Dirac sea that generates
 homogeneous chiral condensates at $\mu=0$.
Exciting the pair of a particle with momentum $\mu$ and a hole with
 momentum $\mu$ ($p=0$) costs zero energy.
Since the pairing wave function $\tan\frac12\theta_p$ at $p>0$ has the
 peak at $p=0$, most likely, the zero energy pairing is generated when
 the interaction is truned on.
This is the Overhauser instability first proposed for spin density wave
 of nonrelativistic electron system \cite{Ov}, and recently revived for
 chiral density wave in QCD \cite{DGR,SS,PRWZ,RSZ}.
Because of effects of this instability, the vacuum must have the chiral
 crystalline structure with a period $\pi/\mu$.

Eq.~(\ref{BV-vac}) tells us an another important result.
Since $\vac$ has nonzero fermion number owing to the existence of the
 Fermi sea and the pairing preserves fermion number, $\vacs$ must also
 have nonzero fermion number.
The expectation value of a fermion number operator becomes
%%%%%%%%%%%%%%%%%
\begin{eqnarray}%
  \langle J^0\rangle &=&
    -\frac12\int\limits_{-\Lambda/2+K}^{\Lambda/2-K}\frac{dp}{2\pi}NL
    +\frac12\int\limits_{-\Lambda/2-K}^{\Lambda/2+K}\frac{dp}{2\pi}NL
    +\int\limits_{\omega_{pa}<0}\frac{dp}{2\pi}NL
    -\int\limits_{\omega_{pb}<0}\frac{dp}{2\pi}NL          \nonumber \\
  &=& \left\{\begin{array}{cc}
        {\displaystyle NL\frac{K}{\pi}}
          & \mbox{for }\mu_\ast^2<\asigma^2 \\
        {\displaystyle NL\left(\frac{K}{\pi}+\frac{\mu_\ast}{\pi}
            \sqrt{1-\frac{\asigma^2}{\mu_\ast^2}}\right)}
          & \mbox{for }\mu_\ast^2>\asigma^2 \\ \end{array}\right. ,
\label{true-FN}
\end{eqnarray}%
%%%%%%%%%%%%%%%
 by our cutoff scheme (\ref{uv-cut}).
At $\mu_\ast=0$, or equivalently $K=\mu$, it yields the correct fermion
 number of the state $\vacs$, $\langle J^0\rangle=NL\mu/\pi$.
We emphasize that the consistent result comes out because of the cutoff
 procedure (\ref{uv-cut}) we employ.
Vanishing fermion number would result if the naive Fourier mode cutoff
 were used.

We comment out about the relation of Eq.~(\ref{true-FN}) to early works
 relating on the zero chemical potential system, finally.
In the chiral sigma model in which $\sigma(x)$ turns into the {\it true}
 backgroun field, the authors of Ref.~\cite{GW} has shown that the
 vacuum expectation value of fermion number density is
 $\langle j^0\rangle=-\frac{\partial_x\varphi}{2\pi}$, where
 $\sigma(x)=\asigma e^{-i\varphi(x)\gamma^5}$, by using the adiabatic
 method.
To compare with our model we set $\varphi(x)=-2Kx$, so that the fermion
 number density becomes the first case of Eq.~(\ref{true-FN}) expecting
 the factor $N$ coming from the number of flavor.
Since the adiabatic method implies the slowly varing condition
 ($\partial_x\varphi\ll\asigma$), the secound case of
 Eq.~(\ref{true-FN}) is excluded from this condition.

%%%%%%%%%%^%%%%%%%%%^%%%%%%%%%^%%%%%%%%%^%%%%%%%%%^%%%%%%%%% Section 5
\section{The Cooper pair model}
\label{sec:CP}

We discuss the Cooper pair model in this section, replacing the
 interaction term in Eq.~(\ref{GN-lag}) with
 $\frac12g^2\abs{\pscb\ps}^2$:
%%%%%%%%%%%%%%%%%
\begin{equation}%
  {\cal L}=\psb\left(i\ds+\gamma^0\mu\right)\ps+\frac12g^2\abs{\pscb\ps}^2.
\label{CM-lag}
\end{equation}%
%%%%%%%%%%%%%%%
The Lagrangian (\ref{CM-lag}) is the same as the one in Ref.~\cite{CMC},
 due to our convention of Dirac matrices.
The Lagrangian has a $U(1)$ symmetry which implies the conserved fermion
 number in the model.
As was shown in Ref.~\cite{CMC} under the assumption of spatilly
 homogenuity, a Lorentz scalar Cooper pair condensate is dynamically
 generated in this model; $\langle\pscb\ps\rangle\neq0$ ($\psch$
 transform as $\ps$ under the Lorentz transformations).

We introduce an auxiliary field $\Delta(x)$ by
%%%%%%%%%%%%%%%%%
\begin{equation}%
  {\cal L} \rightarrow
    {\cal L}-\frac1{2g^2}\abs{\Delta+g^2\pscb\ps}^2.
\end{equation}%
%%%%%%%%%%%%%%%
When one thinks about the Cooper pair condensate, it is convenient to
 regard the charge conjugation of $\ps$ as independent variable of the
 theory (see, for example, Refs.~\cite{CS1,Nambu}).
By introducing the kinetic term of charge conjugation field by
 Eq.~(\ref{kinetic}), we obtain the Hamiltonian density in the following
 form,
%%%%%%%%%%%%%%%%%
\begin{equation}%
  {\cal H}=\frac{\adelta^2}{2g^2}+\frac12\left[\psd,\pscd\right]
    \left[\begin{array}{cc}
      -i\gamma^5\partial_x-\mu & \gamma^0\Delta           \\
       \gamma^0\Delta^\ast     & -i\gamma^5\partial_x+\mu \\
    \end{array}\right]
    \left[\begin{array}{c} \ps \\ \psch \\ \end{array}\right].
\label{hamilton}
\end{equation}%
%%%%%%%%%%%%%%%
As in the case of chiral Gross-Neveu model, we impose the standing wave
 ansatz for the background (auxiliary) field:
%%%%%%%%%%%%%%%%%
\begin{equation}%
  \Delta(x)=\adelta e^{2iKx}.
\end{equation}%
%%%%%%%%%%%%%%%
$\Delta(x)$ is equal to the Cooper pair condensate
 $\langle\pscb\ps\rangle$ apart from a constant factor, so that
 nonvanishing $\adelta$ indicates the symmetry breaking of fermion
 number $U(1)$ symmetry.

Eigenvalues of the first quantized Hamiltonian in Eq.~(\ref{hamilton})
 consist of the following four branches:
%%%%%%%%%%%%%%%%%
\begin{equation}%
\begin{array}{lcl}
  \omega_{pa}=\epsilon_{p_+}+K &,& \omega_{-p\bar{a}}=-\epsilon_{p_-}-K, \\
  \omega_{pb}=\epsilon_{p_-}-K &,& \omega_{-p\bar{b}}=-\epsilon_{p_+}+K, \\
\end{array}
\label{en-spe}
\end{equation}%
%%%%%%%%%%%%%%%
 where $\epsilon_p=\sqrt{p^2+\adelta^2}$ is a energy of a particle with
 momentum $p$ and mass $\adelta$, and $p_\pm=p\mp\mu$ are momenta
 relative to the Fermi surface.
We can then diagonalize the Hamiltonian, and fermionic field is expanded
 in terms of the eigenstates with phase factor, $e^{\pm iKx}$:
%%%%%%%%%%%%%%%%%
\begin{eqnarray}%
  \ps(x)=\int\frac{dp}{2\pi}e^{i(p+K)x}\left[\begin{array}{c}
    \aap\cos\frac12\theta_{p_+}+\bdm\sin\frac12\theta_{p_+} \\
    \bbp\sin\frac12\theta_{p_-}-\adm\cos\frac12\theta_{p_-} \\
  \end{array}\right],                                     \nonumber \\
  \psch(x)=\int\frac{dp}{2\pi}e^{i(p-K)x}\left[\begin{array}{c}
    \bbp\cos\frac12\theta_{p_-}+\adm\sin\frac12\theta_{p_-} \\
    \aap\sin\frac12\theta_{p_+}-\bdm\cos\frac12\theta_{p_+} \\
  \end{array}\right].
\label{field-ex}
\end{eqnarray}%
%%%%%%%%%%%%%%%
The wave functions $\cos\frac12\theta_p$ and $\sin\frac12\theta_p$ are
 defined analogously in Eq.~(\ref{eig-spi}).
The Hamiltonian is obtained in exactly the same form as in
 Eq.~(\ref{ch-hamil}) with the energies of Eq.~(\ref{en-spe}).

Decomposing the fermion number and current operator in the basis of
 Eq.~(\ref{field-ex}) manifests the difference from the chiral
 Gross-Neveu model.
In the Cooper pair model, fermion current is diagonalized by the
 quasi-particle operators as the following equation, but fermion number
 is not;
%%%%%%%%%%%%%%%%%
\begin{eqnarray}%
  J^1 &=& \frac12\int dx\left[
            \psd\gamma^5\ps-\pscd\gamma^5\psch\right]          \nonumber \\
      &=& \frac12\int\frac{dp}{2\pi}\left(
            \adp\aap-\bdp\bbp-\aam\adm+\bbm\bdm \right).
\end{eqnarray}%
%%%%%%%%%%%%%%%

The cutoff scheme is the same one as the chiral Gross-Neveu model,
 because the field expansion~(\ref{field-ex}) must reduce to the free
 theory even if $\adelta=\mu=0$ and $K\neq0$ (see Appendix).
Variational ground state $\vacd$ is defined by filling the negative
 energy states in the same way as Eq.~(\ref{sig-vac}), and then the
 effective potential becomes
%%%%%%%%%%%%%%%%%
\begin{eqnarray}%
  {\cal V}(\adelta,K;\mu) &=& \frac{\langle H\rangle}{NL} \nonumber \\
    &=& \frac{\adelta^2}{2\lambda_0}
       +\frac12\int\limits_{-\Lambda/2+K}^{\Lambda/2-K}
          \frac{dp}{2\pi}\omega_{-p\bar{a}}
       +\frac12\int\limits_{-\Lambda/2-K}^{\Lambda/2+K}
          \frac{dp}{2\pi}\omega_{-p\bar{b}}               \nonumber \\
    & & \hspace{30pt}
       +\int\limits_{\omega_{pa}<0}\frac{dp}{2\pi}\omega_{pa}
       +\int\limits_{\omega_{pb}<0}\frac{dp}{2\pi}\omega_{pb}.
\label{copo}
\end{eqnarray}%
%%%%%%%%%%%%%%%
In (\ref{copo}), $\langle H\rangle$ implies to take expectation value
 with respect to the state $\vacd$.
Performing the integrations explicitly and using the renormalization
 condition (\ref{ren-cond}) (replacing $\asigma$ to $\adelta$ in the
 equation), we obtain the renormalized effective potential as
%%%%%%%%%%%%%%%%%
\begin{eqnarray}%
  K^2<\adelta^2 &:& {\cal V}=\frac{\adelta^2}{4\pi}\left[
     \frac{2\pi}{\lambda}-3+\ln\frac{\adelta^2}{m_0^2}\right]
    +\frac{K^2}{2\pi}-\frac{\mu^2}{2\pi},                  \nonumber \\
  K^2>\adelta^2 &:& {\cal V}=\frac{\adelta^2}{4\pi}\left\{
     \frac{2\pi}{\lambda}-3+\ln\frac{K^2}{m_0^2}\left[
       1+\sqrt{1-\frac{\adelta^2}{K^2}}\right]^2\right\}             \\
  && \hspace{144pt}
    +\frac{K^2}{2\pi}\left[1-\sqrt{1-
       \frac{\adelta^2}{K^2}}\right]-\frac{\mu^2}{2\pi}.   \nonumber
\end{eqnarray}%
%%%%%%%%%%%%%%%
It is easy to observe that the effective potential is identical with
 that of the chiral Gross-Neveu model, Eq.~(\ref{R-poten}), under the
 replacement of $\mu_\ast$ by $K$.
Then, the vacuum is obtained at $K=0$, and the Cooper pair condensate of
 vacuum is given by
%%%%%%%%%%%%%%%%%
\begin{equation}%
  \Delta(x)=m_0e^{1-\pi/\lambda}=-\lambda_0\langle\pscb\ps\rangle/N
\end{equation}%
%%%%%%%%%%%%%%%
(The last equality is straightforwardly proved, similarly to
 Eq.~(\ref{chi-cond}) in the chiral Gross-Neveu model).
Therefore, the ground state with spatially uniform Cooper pair
 condensate obtained in Ref.~\cite{CMC} gives the true vacuum in this
 model.

The question we have to ask here is why the Cooper pair condensate is
 uniform differently from the chiral condensate in the chiral
 Gross-Neveu model.
In order to answer this question, we had to better examine the
 fermion number and current of the variational ground state $\vacd$.
After some calculations, they turn out to be
%%%%%%%%%%%%%%%%%
\begin{eqnarray}%
  \langle J^0\rangle &=& NL\frac{\mu}{\pi}, \\
  \langle J^1\rangle
    &=& \left\{\begin{array}{cc}
          {\displaystyle NL\frac{K}{\pi}}
            & \mbox{for }K^2<\adelta^2 \\
          {\displaystyle NL\left(\frac{K}{\pi}-\frac{K}{\pi}
              \sqrt{1-\frac{\adelta^2}{K^2}}\right)}
            & \mbox{for }K^2>\adelta^2 \\ \end{array}\right. .
\label{J^1}
\end{eqnarray}%
%%%%%%%%%%%%%%%
The fermion number is independent of the parameter indicating the
 spatial varidity $K$ in constrast to the chiral Gross-Neveu model.
The spatial homogeneity can be explicitly broken in the chiral
 Gross-Neveu model because the fermion number is an order parameter for
 this symmetry breaking and it has a source, namely chemical potential.
In the Cooper pair model this symmetry breakdown, if it occurs, must be
 caused spontaneously because the fermion number is no longer the order
 parameter.
Then, the spontaneous symmetry breaking of the spatial homogeneity does
 not occur in our models.
%The proposition is not applicable to the case when Cooper pairs with a
% nonvanishing total momentum condense, as seen in
% Ref.~\cite{ABR,BKRS,LRS}.
%If we live in the rest frame of the Cooper pair, background fermion
% induced by chemical potential should flow.
%In other words, there is fermion current source in this frame.
%It is, then, expected that $K$ has a nonzero value such that
% $\langle J^1\rangle$ becomes finite.
%%% replaced by

%By the similar calculation in this paper, we can show that crystalline
% Cooper pair condensate appears for the model with fermion current
% source, $\rho\,\psb\gamma^1\ps$.
%The statement relates to the QCD with a mismatch in the fermi surface of 
% up and down quarks in Refs.~\cite{ABR,BKRS,LRS}, because the Cooper
% pairs with a nonvanishing total momentum for the QCD model imply the
% flow of fermion current.

%%%%%%%%%%^%%%%%%%%%^%%%%%%%%%^%%%%%%%%%^%%%%%%%%%^%%%%%%%%% Section 6
\section{Conclusions}
\label{sec:conc}

In this paper we have analyzed the vacuum structure of (1+1)-dimensional
 models with quartic fermi interactions at finite densities, with
 particular emphasis on the possibility of spatially nonuniform ground
 state.
Our treatment is restricted into the case of large-$N$ limit and of zero
 temperature. This is the unique place where we can have nonvanishing
 order parameter in 1+1 dimensions in the sense explained in
 Introduction.

We first examined the chiral Gross-Neveu model.
We have constructed the effective potential by allowing the chiral
 condensate to vary periodically in space.
The quasi-particle basis diagonalizing the fermionic Hamiltonian of the
 theory was obtained by the Fourier expansion after performing the
 chiral rotation.
We have constructed an effective potential and have shown that the
 vacuum has a crystalline structure in which the chiral condensate
 oscillates with the wave number $2\mu$.
This is consistent with the result obtained by Sch\"{o}n and Thies
 \cite{ST}.

In the course of constructing the effective potential, we have found
 that the branch-dependent ultraviolet cutoff procedure is required to
 derive the correct effective potential.
In the cutoff scheme, the shifts from the explicit cutoff parameter play
 important role for the purpose of providing the system with the correct
 fermion number.
While the branch-dependent ultraviolet cutoff procedure may look ugly,
 it is likely that it is required to preserve the fermion number in the
 system by cancelling the spectral flow induced by chiral rotation
 performed to obtain the basis used in our treatment.

We have also constructed the variational ground state in terms of free
 particle basis and performed the Bogoliubov-Valatin transformation to
 obtain quasi-particle basis which diagonalize the Hamiltonian.
We have shown that the wave number $2\mu$ of the chiral condensate
 results from the Overhauser instability which generates
 particle-hole/antiparticle pairs with the total momentum $2\mu$, or
 $-2\mu$.

We have examined the possibility of spatially varying ground state in
 the Cooper pair model proposed by Chodos, et al.~\cite{CMC} by allowing
 the Cooper pair condensate to vary periodically in space.
By constructing and analyzing the effective potential, we have shown
 that the ground state with spatially uniform Cooper pair condensate
 obtained in Ref.~\cite{CMC} gives the true ground state of this model.

%%% added!
We should note that our model, which contains a single chemical
 potential common to $N$-flavor fermions, does not quite mimic the QCD
 motivated (3+1)-dimensional model in Refs.~\cite{ABR,BKRS,LRS} in which
 the mismatch in the Fermi surfaces of up and down quarkes are the
 essential ingredient for the crystalline color superconducting phase.
It would be very interesting to construct a (1+1)-dimensional model
 which possesses the analogous structure and work it out in the similar
 way as done in this paper. However, the formulation of the large-$N$
 field theory with two different chemical potentials is highly
 nontrivial and we did not enter into the problem in this paper.
%%%

The spatial variations of Cooper pair condensate, if it forms, is due to
 the condensation of particle-particle pairs (or
 antiparticle-antiparticle pairs) with nonvanishing total momenta, as we
 observe from the Bogoliubov-Valatin transformation.
It means that there is nozero expectation value of fermionic current.
In other words, the fermionic current becomes an order parameter for the
 breakdown of the spatial homogeneity.

On the other hand, the role of order parameter is played by fermion
 number in the chiral Gross-Neveu model, so that the breakdown of the
 spatial homogeneity is caused explicitly by the chemical potential (in
 fact, the symmetry is restored at $\mu=0$).
In the Cooper pair model, however, this symmetry breakdown is caused
 spontaneously if it occurs.
The result, therefore, shows that the spatial homogeneity is not
 sopntaneously broken in our models.

%%%%%%%%%%^%%%%%%%%%^%%%%%%%%%^%%%%%%%%%^%%%%%%%%%^%%%%%%%%% Acknowlegements
\section*{Acknowledgements}

I wish to thank Professor W. A. Bardeen for suggesting possible
 connection between the regularization scheme employed in this paper and
 the spectral flow due to chiral rotation in the context of chiral
 anomaly.
I also thank Professor H. Minakata for suggesting the problem and for
 careful reading of this paper.

%%%%%%%%%%^%%%%%%%%%^%%%%%%%%%^%%%%%%%%%^%%%%%%%%%^%%%%%%%%% Appendix
\appendix
\section{Ultraviolet regularization}
\label{sec:UV}

In this Appendix we explain the ultraviolet cutoff procedure given in
 Eq.~(\ref{uv-cut}) in detail.

If we set $\asigma=0$ in Eq.~(\ref{sigma-ans}) in the chiral Gross-Neveu
 model, it becomes a free field theory.
In this case the quasi-particle description should reduce to the
 free field description.
Eq.~(\ref{mode-ex}), however, differs from Eq.~(\ref{free-ex}) by the
 factor $e^{\pm iKx\gamma^5}$.
We will show below that the superficial difference between the
 quasi-particle and the free-particle description is removed by the
 ultraviolet cutoff prescription.

From the Bogoliubov transformation (\ref{BV-trans}) at $\asigma=0$ we
 first obtain
%%%%%%%%%%%%%%%%%
\begin{equation}%
\begin{array}{lcc}
  a_p=a_{p+K}^{(0)} & \mbox{ for} & p>0, \\
  a_p=a_{p-K}^{(0)} & \mbox{ for} & p<0. \\
\end{array}
\label{BV1}
\end{equation}%
%%%%%%%%%%%%%%%
It is natural that the free field description possesses naive momentum
 cutoff, $\abs{q}<\Lambda/2$ in Eq.~(\ref{free-ex}).
We observe from Eq.~(\ref{BV1}) that the cutoff in the $a$-branch
 becomes $-\Lambda/2<p-K=q<-K$ and $K<q=p+K<\Lambda/2$, i.e.
 $\abs{p}<\Lambda/2-K$.
The remaining mode of free fermion, $-K<q<K$, comes from the
 $b$-branch:
%%%%%%%%%%%%%%%%%
\begin{equation}%
\begin{array}{lcc}
  b_p= a_{-p+K}^{(0)\dag} & \mbox{ for} &  0<p<K, \\
  b_p=-a_{-p-K}^{(0)\dag} & \mbox{ for} & -K<p<0. \\
\end{array}
\end{equation}%
%%%%%%%%%%%%%%%
Finally, the rest of the Bogoliubov transformation,
%%%%%%%%%%%%%%%%%
\begin{equation}%
\begin{array}{lcc}
  b_p=b_{p-K}^{(0)} & \mbox{ for} & p> K, \\
  b_p=b_{p+K}^{(0)} & \mbox{ for} & p<-K, \\
\end{array}
\label{BV3}
\end{equation}%
%%%%%%%%%%%%%%%
 leads to the cutoff in the $b$-branch, $-\Lambda/2<p+K=q$ and
 $q=p-K<\Lambda/2$, namely $\abs{p}<\Lambda/2+K$.
In addition, we can verify by using Eqs.~(\ref{BV1})--(\ref{BV3}) that
 the quasi-particle decomposition (\ref{mode-ex}) with the cutoff
 (\ref{uv-cut}) is reduced to the free-particle decomposition
 (\ref{free-ex}) with the naive momentum cutoff, $\abs{q}<\Lambda/2$.

In the above discussion, we have proved that the ultraviolet cutoff must
 be Eq.~(\ref{uv-cut}) for $\asigma=0$ at least.
We believe that this cutoff is correct one at $\asigma\neq0$ because the
 cutoff procedure must be insensitive to the vacuum expectation value,
 as mentioned in Sec.~\ref{sec:GN}; thus, this cutoff is required to
 provide the state $\vacs$ with the correct fermion number.

\begin{figure}
\begin{center}
 \epsfig{file=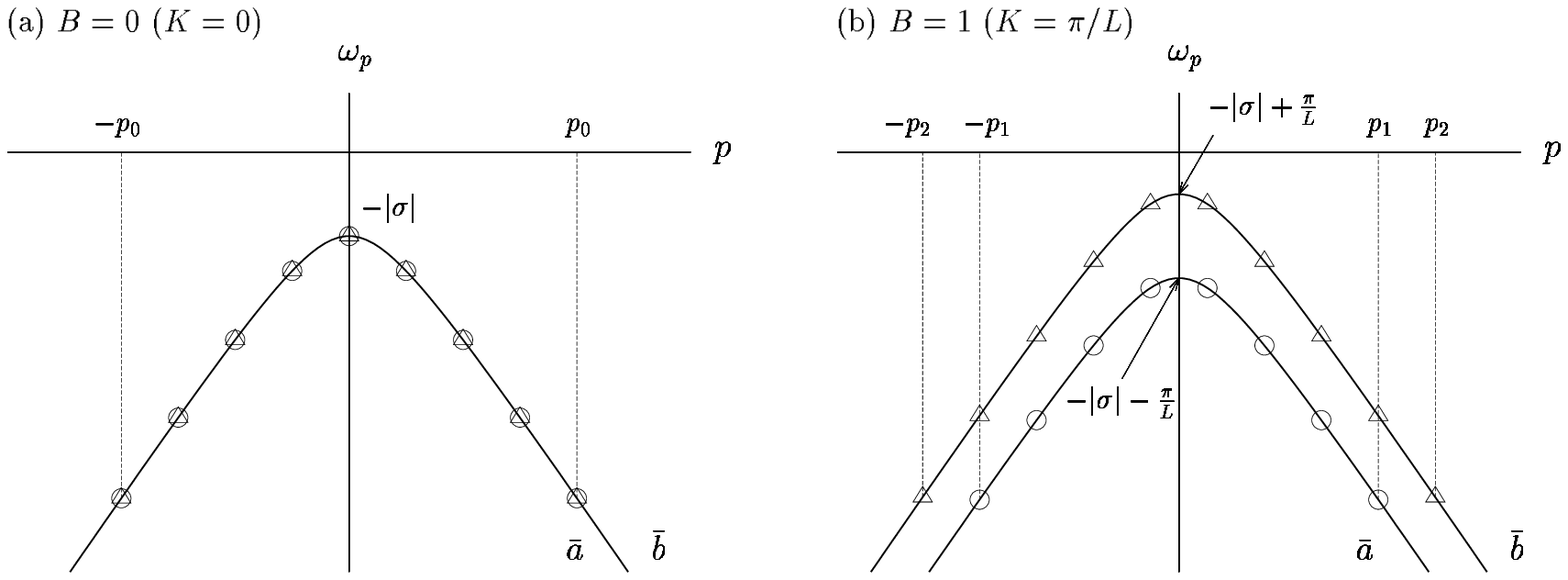,width=\linewidth}
\end{center}
\caption{The cutoff scheme at $\mu=0$.
 The effective potential is given by the sum of the energies at circles
 ($\bar a$-branch) and triangles ($\bar b$-branch).
 The parameters in the graphs are
 $p_0=\frac{2\pi}{L}N_\Lambda=\frac{\Lambda}{2}$,
 $p_1=\frac{2\pi}{L}\left(N_\Lambda-\frac12\right)
     =\frac{\Lambda}{2}-\frac{\pi}{L}$ and
 $p_2=\frac{2\pi}{L}\left(N_\Lambda+\frac12\right)
     =\frac{\Lambda}{2}+\frac{\pi}{L}$, respectively. }
\label{fig:cutoff}
\end{figure}

Now, we comment on the cutoff prescription from a different point of
 view.
We enclose the system in a finite box of length $L$ for the purpose of
 counting the number of modes correctly.
We shall impose periodic boundary conditions on $\ps(x)$ and $\psch(x)$.
This leads to the following conditions on $K$ and $p$ because of the
 chiral phase factor in wave functions (\ref{eigst}),
%%%%%%%%%%%%%%%%%
\begin{equation}%
  K=\frac{\pi B}{L},\;\;\;p=\frac{2\pi}{L}\left(n+\frac12B\right),
\end{equation}%
%%%%%%%%%%%%%%%
 where $B$ and $n$ are integer.
One can see from the cutoff (\ref{uv-cut}) that $n$ take
 $(2N_\Lambda-B+1)$ values from $-N_\Lambda$ to $N_\Lambda-B$ for
 $\bar{a}$-branch, and $(2N_\Lambda+B+1)$ values from $-N_\Lambda-B$ to
 $N_\Lambda$ for $\bar{b}$-branch, respectively, for fixed $B$ (i.e.,
 fixed $K$).
Here, $N_\Lambda=\Lambda L/4\pi$ is an explicit cutoff parameter.
The number of modes included in divergent sums to evaluate the
 effective potential (\ref{effec}) is kept constant, independently of
 $B$; $(2N_\Lambda-B+1)+(2N_\Lambda+B+1)=2(2N_\Lambda+1)$.
We depict the cutoff prescription in Fig.~\ref{fig:cutoff}.
This figure tells us that the cutoff implies summing modes in order from
 above with respect to the absolute value of energy, keeping the total
 number of modes fixed when $\mu=0$.

We also note that the fermion number of the state $\vacs$ results from
 the difference of mode between $\bar{a}$- and $\bar{b}$-branch when we
 take the cutoff as Fig.~\ref{fig:cutoff}.
Since $\bar{a}$-branch carries fermion number $-\frac12$ and
 $\bar{b}$-branch carries $+\frac12$ as one can find from
 Eq.~(\ref{charge}), the fermion number of $\vacs$ becomes
 $-\frac12(2N_\Lambda-B+1)+\frac12(2N_\Lambda+B+1)=B$ times $N$ of
 flavor, namely, $NL\frac{K}{\pi}$.
It is consistent with the first equation of Eq.~(\ref{true-FN}) because
 we now ignore the region where $\omega_{-p\bar{a},\bar{b}}>0$.

For the Cooper pair model, we can discuss the cutoff procedure as the
 chiral Gross-Neveu model.
By comparing Eq.~(\ref{field-ex}) to Eq.~(\ref{free-ex}), we obtain the
 Bogoliubov transformation
%%%%%%%%%%%%%%%%%
\begin{eqnarray}%
   K<p    &:& \left\{\begin{array}{l}
                a_p = a_{ p+K}^{(0)    }\cos\frac12\theta_{p_+}
                     -a_{-p+K}^{(0)\dag}\sin\frac12\theta_{p_+},\\
                b_p = b_{ p-K}^{(0)    }\cos\frac12\theta_{p_-}
                     -b_{-p-K}^{(0)\dag}\sin\frac12\theta_{p_-},\\
              \end{array}\right. \nonumber        \\
  -K<p< K &:& \left\{\begin{array}{l}
                a_p = a_{ p+K}^{(0)    }\cos\frac12\theta_{p_+}
                     +b_{ p-K}^{(0)    }\sin\frac12\theta_{p_+},\\
                b_p = a_{-p+K}^{(0)\dag}\cos\frac12\theta_{p_-}
                     -b_{-p-K}^{(0)\dag}\sin\frac12\theta_{p_-},\\
              \end{array}\right. \label{CP-Btrans}\\
     p<-K &:& \left\{\begin{array}{l}
                a_p = b_{-p-K}^{(0)\dag}\cos\frac12\theta_{p_+}
                     +b_{ p-K}^{(0)    }\sin\frac12\theta_{p_+},\\
                b_p = a_{-p+K}^{(0)\dag}\cos\frac12\theta_{p_-}
                     +a_{ p+K}^{(0)    }\sin\frac12\theta_{p_-}.\\
              \end{array}\right. \nonumber
\end{eqnarray}%
%%%%%%%%%%%%%%%
The condition in which the quasi-particle description is reduced to the
 interaction-free description at $\adelta=0$ in spite of the presence of
 $K$, leads to the cutoff (\ref{uv-cut}).

We also depict the same figure as Fig.~\ref{fig:cutoff} because the
 dispersion relations of quasi-particles in the two models coinside with
 each other at $\mu=0$.
In the Cooper pair model, however, the quasi-particle does not carry
 definite fermion number, but the fermion current.
Therefore, the fermion current of the state $\vacd$ is generated as we
 vary the value of $K$, as indicated in Eq.~(\ref{J^1}).
This means from the viewpoint of the Bogoliubov transformation
 (\ref{CP-Btrans}) that $\vacd$ is composed of two (free)fremions
 pairing with total momentum $2K$ or two antifermions pairing with total
 momentum $-2K$.

%%%%%%%%%%^%%%%%%%%%^%%%%%%%%%^%%%%%%%%%^%%%%%%%%%^%%%%%%%%% References

%%%%%%%%%%^%%%%%%%%%^%%%%%%%%%^%%%%%%%%%^%%%%%%%%%^%%%%%%%%% Figures

\vspace{36pt}

\end{document}